\documentstyle[12pt,a4p,12pt,cite,epsfig]{article}
\def\RI{$R_{1\rm{D}}\ $}

\def\NJK{\langle N_{\rm{jet}}(k) \rangle\ }
\def\NJM{\langle N_{\rm{jet}}(m) \rangle\ }
\def\NJKI{\langle N_{\rm{jet}}(\sqrt{s_k}) \rangle\ }
\def\NJMI{\langle N_{\rm{jet}}(\sqrt{s_m}) \rangle\ }

\def\Nch{$N_{\rm{ch}}$}
\def\nch{N_{\rm{ch}}}

\def\z0{${\rm Z}^0$}

\begin{document}
\begin{titlepage}
\begin{flushright}
TAUP-2937/11
\end{flushright}
\vspace{7mm}

\begin{center}
\vspace{8mm}
{\Large\it The dependence of Bose-Einstein Correlations\\ 
\vspace{2mm}

on energy, multiplicity and hadronic jets}
 
\vspace{1mm}
\end{center}
\bigskip
\begin{center}   
\vspace{8mm}   
{\large\bf Gideon Alexander\footnote{Email: \it gideona@post.tau.ac.il}}
\end{center}  
\vspace{2mm}
\centering{\it School of Physics and Astronomy}\\
\centering{\it Raymond and Beverly Sackler Faculty of Exact Sciences}\\
\centering{\it Tel-Aviv University, 69978 Tel-Aviv, Israel}\\

\vspace{10mm}
  
\vspace{5mm}
\begin{flushleft}

\begin{abstract} 
{\small The dependence of the one dimensional Bose-Einstein correlation
\RI derived from  proton-proton collisions is examined in
terms of energy, hadron multiplicity and number of emerging
hadronic jets. It is argued that the observed rise of
\RI with energy is the result of the increase
of the average number of hadronic jets. It is further shown that
by the relative straightforward measurements of \RI and its chaoticity 
parameter $\lambda$ one should be able to
estimate the average number of outgoing hadronic jets emitted over
the entire final state phase space without the 
necessity to rely on a particular jet finding algorithm and its
associated parameters.  
}
\end{abstract}
\end{flushleft}

\vspace{1cm}
\today
\vspace{1cm}
\begin{flushleft}
{\small \it PACS:} {\small 13.75,Cs;13.85.Hd;25.75.Gz}\\
{\it Keywords}: {\small Bose-Einstein correlations;
energy; charge multiplicity; hadronic jets}
\end{flushleft}

\end{titlepage}
\vspace{5mm}
\begin{flushleft}

\section{Introduction}
The Bose-Einstein Correlation (BEC) of identical
boson pairs, mainly $\pi^{\pm}\pi^{\pm}$ mesons, 
produced in particles and heavy ion collisions
has been studied since the early 1960's
\cite{goldhaber}. In the one-dimensional BEC analyzes one 
frequently utilizes
the Lorentz invariant variable $Q^2$ which is defined as
$Q^2 =-(\tilde{p_1}-\tilde{p_2})^2$ where $\tilde{p}_{i}$ is
the four momentum of the $ith$ identical boson. Taking the emitting source
of the bosons to be described by a spherical symmetric Gaussian
density distribution, the correlation function of two
identical bosons can then be given e.g. by
\begin{equation}
C(Q)\ =\ N(1\ +\ \lambda e^{-R_{\rm{1D}}^2Q^2})(1+\delta Q +\epsilon Q^2)\ ,
\label{eq_main}
\end{equation}
where \RI (also referred to at times as $R_{\rm{G}}$ or $R_{\rm{inv}}$) is the one
dimensional (1D) 
value of the particles' emitter source and
$\lambda$ is the strength of the BEC effect. This $\lambda$,
which is the chaoticity parameter, lies
within the range of $0 \leq \lambda \leq 1$ \cite{review,kittel}.
The parameters $\delta\ {\rm{and}}\ \epsilon$ are frequently
introduced to account
for possible long range correlations, 
like those related to momentum
and energy conservations, while $N$ is a normalization factor. 
The results obtained from BEC analyzes depend among others 
on the assumption that the low percentage of the non pion outgoing charged
tracks does not influence the BEC results. Furthermore 
the particular choice taken from the various reference
samples against which the BEC effect is compared to 
has to be kept in mind when common underlying features are searched for 
in the BEC analyzes results. In addition  
the reported \RI values for charged identical bosons
were not always a subject to a   
Coulomb correction which however is small and estimated to be $\leq
0.05$ fm 
( see e.g. Ref. \cite{alice_09}). In more
recent studies the BEC analyzes
were extended to two and three dimensions to allow for
non-spherical
emitting sources \cite{review}.
\vspace{2mm}

Our present study
concentrates on the 1D case which allows us to 
include early BEC analyzes of $pp\to hadrons$ collisions 
carried out at relatively low energies. The main aim of 
the present work
is to study the \RI
dependence on the $pp$ colliding center of mass energy 
$\sqrt{s}$, the outgoing 
charged hadron multiplicity and hadronic jets and 
to explore the possibility
to use it for the estimation of the  
average number of 
hadronic jets emitted over the full final state phase space.

\section{The \RI dependence on the $pp$ colliding energy}  
The recent operation of the 
Large Hadron $pp$ Collider (LHC) at CERN has opened up 
the opportunity to examine the \RI derived from BEC 
of identical pion pairs produced in $pp$ collisions at very high
center of mass energies   
up to 7 TeV . 
A set of \RI values, obtained via Eq. (\ref{eq_main}) and taken from
Refs \cite{alice_09,e766,na23,na27,cms_09_7,cms_236}
are presented in Fig. \ref{rpp} as
\begin{figure}[h]
\centering{\epsfig{file=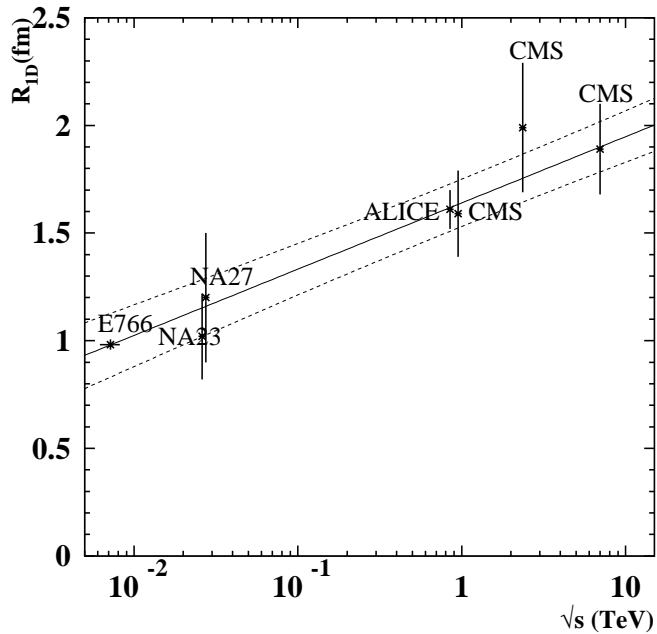,height=12cm,bbllx=36pt,
bblly=150pt,bburx=554pt,bbury=879pt,clip=}}
\caption{\small A set of \RI values 
and their errors obtained from BEC analyzes, using 
Eq. (\ref{eq_main}), of identical charged
pion pairs produced in $pp$ collisions at several
center of mass energies.
\cite{alice_09,e766,na23,na27,cms_236,cms_09_7}
averaged over all charge multiplicities.                                       
The middle line represents
$R_{\rm{1D}}=1.64+0.14\times \ln(\sqrt{s})$ fm as obtained
from a fit to the data. The dashed lines define the $\pm$1 s.d.
error band associated with the fitted parameters. 
}
\label{rpp}

\end{figure}
a function of the $pp$ colliding energy in
the range of $\sqrt{s}$ from 5 GeV to 15 TeV . 
The errors shown 
are the statistical and systematic ones added in 
quadrature. From the two ALICE values at 0.9 TeV reported 
in Ref. \cite{alice_09}, the one
with the considerably smaller systematic error was chosen.
In the figure the CMS and ALICE values at 0.9 TeV
are for clarity shifted slightly apart. 
Here it should be noted that the data selection criteria 
used for the \RI measurements at the LHC energies
are somewhat different from those used
in the BEC analyzes at the lower energies.
In particular at the LHC energies 
the observed
hadron tracks used for the correlation analyzes are confined 
to a restricted 
phase space domain commonly specified in terms of the 
experimentally accessible
pseudorapidity $\eta$ range where $\eta$ is defined as
$$\eta\ =\ 0.5\times \ln[(|\vec{p}|+p_L)/(|\vec{p}|-p_L)].$$
Here $|\vec{p}|$ is the absolute value of the
outgoing particle  
three momentum vector and $p_L$ is its longitudinal 
momentum component. However it has already been previously observed
\cite{bec_rapidity} that the BEC lengths show, if at all, 
only a weak dependence on $\eta$.
\vspace{2mm}

A clear rise of \RI with $\sqrt{s}$ is seen in Fig. \ref{rpp}
and is here represented by the continuous line which
was obtained from a fit of the expression
\begin{equation} 
R_{\rm{1D}}\ =\ a\ +\ b\times \ln(\sqrt{s})\ \rm{fm}
\label{r1d}
\end{equation}
to the data yielding $a=1.64\pm 0.11$ fm
and $b=0.14\pm 0.02$ fm with a $\chi^2/dof$ of 0.3/5 where $\sqrt{s}$
is given in TeV units. The deviation of $\pm$ 1 s.d. is indicated by
the dashed lines in the same figure.
At first one is lead to presume that the \RI increase  
is a direct consequence of the energy rise however, 
this apparently cannot be the prime reason as
has been recently clearly demonstrated by the 
CMS collaboration \cite{cms_09_7}. 
In their BEC study they have
compared the \RI values measured in $pp$
collisions at the two energies of
0.9 and 7.0 TeV as a function of the outgoing
charge multiplicity \Nch which are
shown in Fig. \ref{cms_energy}.
The dotted and solid curves in the figure are the results from 
the fits of the  
relation \cite{lisa} $R_{\rm{1D}}(\nch)=k\times {\nch}^{1/3}$ to the data
yielding the values $k=0.597\pm 0.009 \pm 0.057$ fm 
and $k= 0.612\pm 0.007\pm 0.068$ fm 
at 0.9 and at 7.0 TeV  respectively.
\begin{figure}[ht]
\centering{\epsfig{file=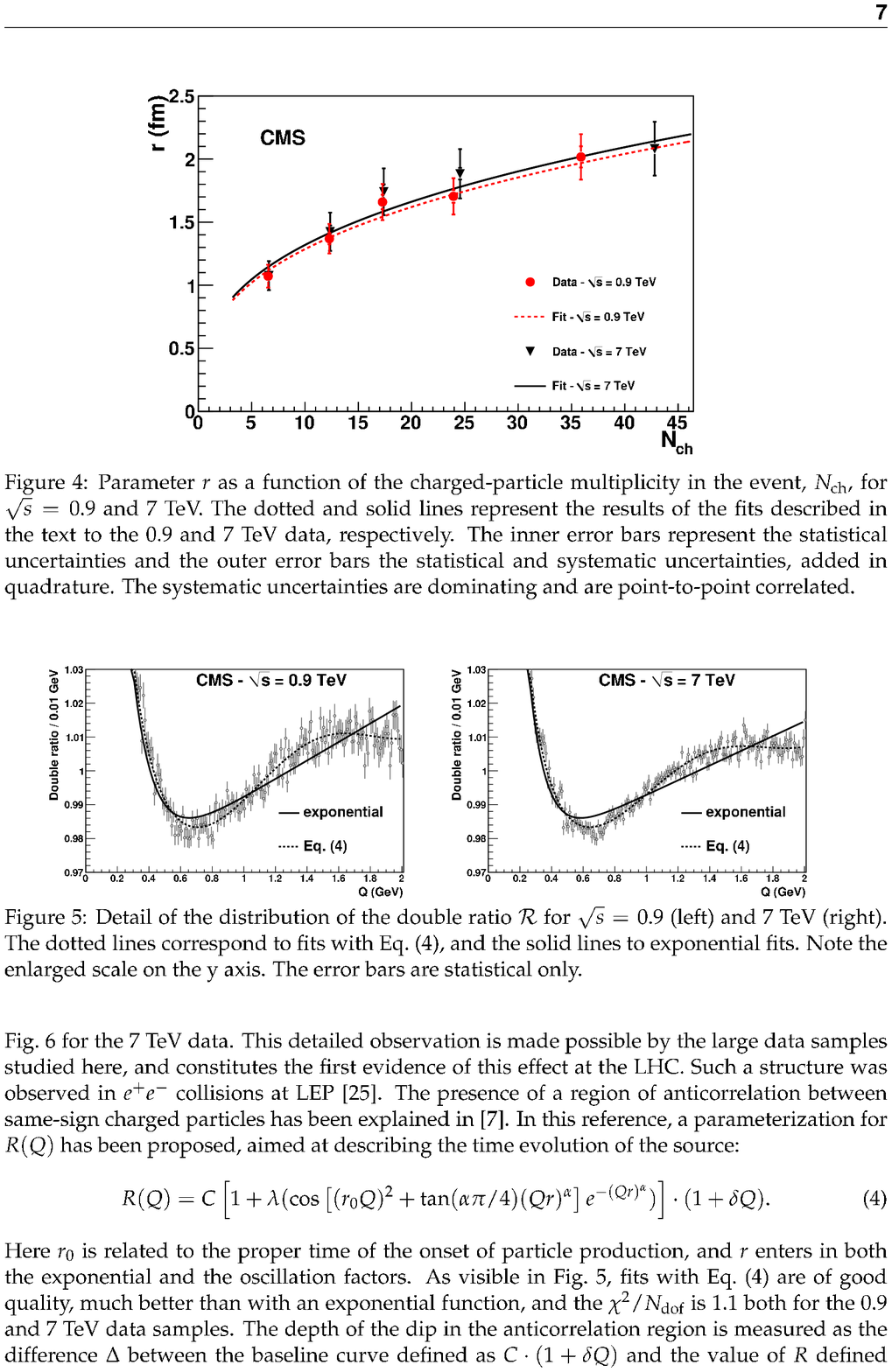,
height=6.5cm,bbllx=110pt,bblly=529pt,bburx=550pt,bbury=737pt,clip=}}
\caption{\small The \RI values measured by the CMS 
collaboration \cite{cms_09_7}
as a function of the charged-particle multiplicity $Nch$ at
$\sqrt{s}$ = 0.9 and 7.0 TeV . The dotted and solid lines represent the
fit results of the relation $R_{\rm{1D}}(\nch)=k\times \nch^{1/3}$
to the 0.9 and 7.0 TeV data respectively. The inner error bars
represent the statistical uncertainties
and the outer error bars the statistical and systematic uncertainties
added in quadrature. 
}
\label{cms_energy}
\end{figure}
\begin{figure}[ht]
\centering{{\epsfig{file=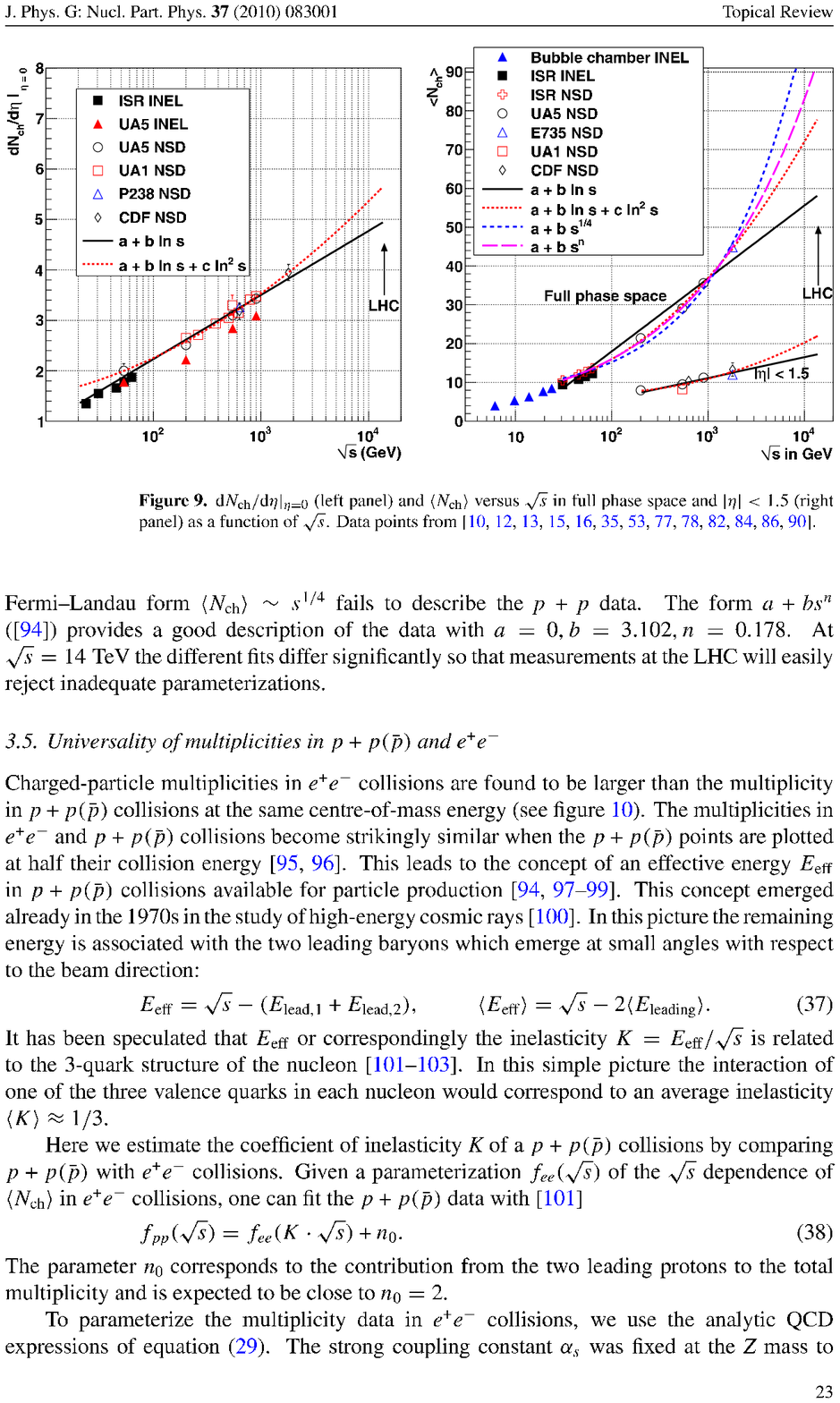,height=7.3cm,
bbllx=280pt,bblly=556pt,bburx=474pt,bbury=751pt,clip=}}  
{\epsfig{file=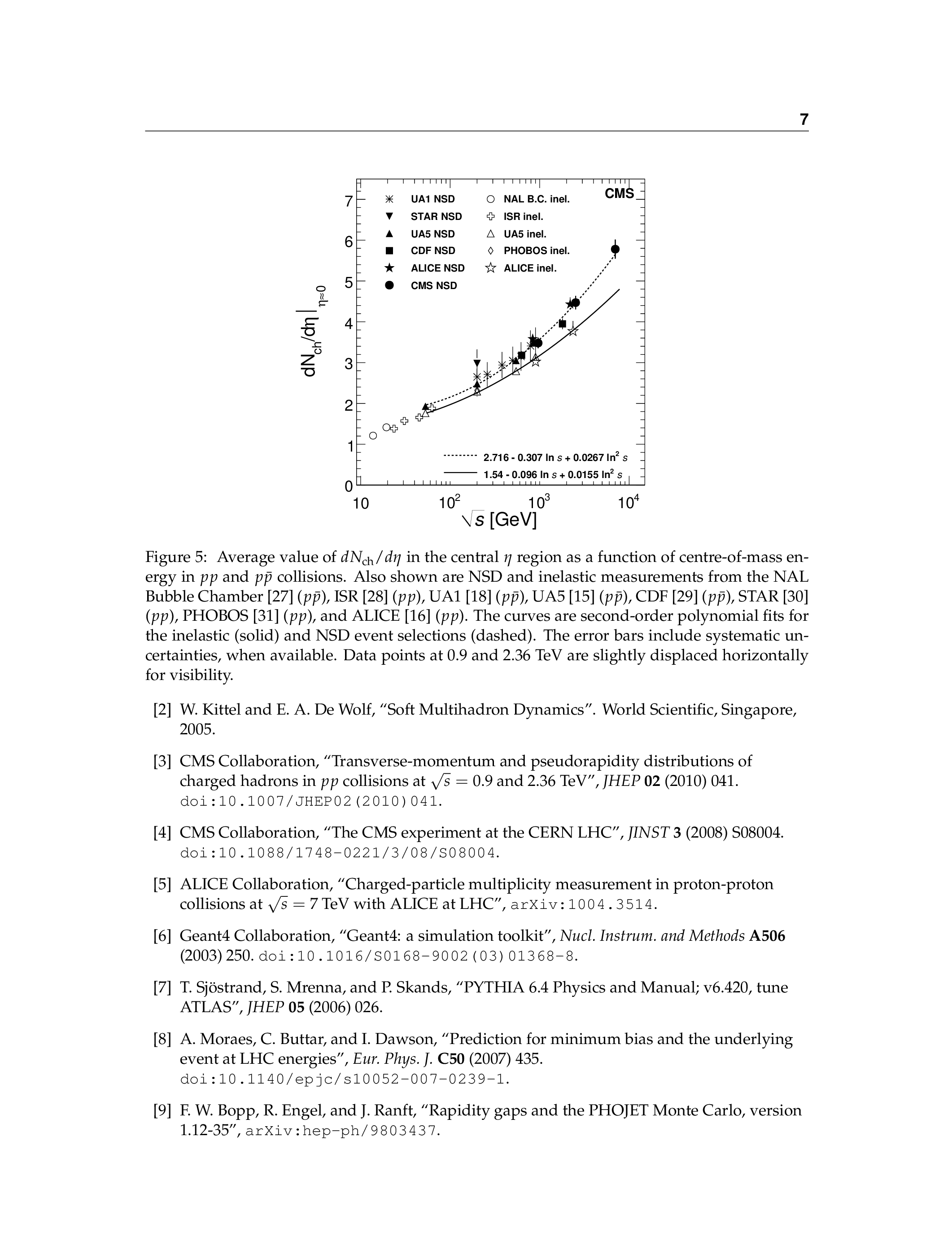,height=7.3cm,
bbllx=160pt,bblly=450pt,bburx=430pt,bbury=693pt,clip=}}}
\caption{\small 
Left: A compilation, taken from \cite{multi_review},
of the average charge multiplicity $\langle$\Nch $\rangle$ 
produced in $pp$ and $\bar pp$ reactions versus 
$\sqrt{s}$ covering the full phase space and
within $|\eta| \leq$1.5; 
Right: A compilation, taken from \cite{cms7} of $d\nch/d\eta$,
versus $\sqrt{s}$ obtained from from $pp$ and $\bar pp$ inelastic (INEL) 
reactions within $|\eta| \leq 0.5$ and non$-$single diffractive 
collisions (NSD) within  $|\eta| \leq 1$. 
The lines shown in 
the figures represent the fits to a power series in $\ln{s}$.  
}
\label{mult1}
\end{figure} 
As can be seen, the two fitted curves almost coincide so that 
one can safely conclude
that \RI is essentially independent of the energy but is
correlated to the increase in the 
charge multiplicity as has already 
been observed 
in the studies of $pp$ and light ion collisions  
\cite{barshay} as well as in the $e^+e^- \to Z^0 \to hadrons$ 
annihilations \cite{opal_mult3}. 

\section{The average charge multiplicity as a function of energy}
The average final state charge multiplicity $\langle \nch \rangle$
produced in $pp$
and $\bar pp$ collisions is known to increase with
energy as shown in  
the left part of Fig. \ref{mult1} which was 
taken from Ref. \cite{multi_review}.
The $\langle \nch \rangle$ values shown were obtained from
the charged tracks emitted in 
$pp$ and $\bar pp$ collisions summed up over the entire phase space 
which was experimentally quite accessible in reactions  
at a relatively low $\sqrt{s}$ regions.
However, it became more difficult as 
the energy increased with the hadron colliders 
to the values near 2 TeV at the
$\bar pp$ Tevatron accelerator and even higher to
the current value of 7\ TeV supplied by
the LHC at CERN where the analysis was
restricted to a limited phase space 
defined by a range in the pseudorapidity $\eta$.
In this limited phase space  
one has still the possibility to determine $d\nch/d\eta$ i.e., the 
differential charge multiplicity
density as a function of $\eta$.
To represent a $d\nch/d\eta$  distribution one often uses 
its specific value at $\eta=0$ that is $d\nch/d\eta|_{\eta=0}$. 
The right part of Fig. \ref{mult1}, taken from reference
\cite{cms7}, shows the 
charge multiplicity density at  $\eta$=0  
obtained in $pp$ and $\bar pp$ collisions as a 
function of $\sqrt{s}$. A clear rise of both $\langle \nch \rangle$
 and $d\nch/d\eta|_{\eta=0}$ with $\sqrt{s}$ is seen 
which can be parametrized by a power series in $\ln(s)$ 
as indicated in Fig. \ref{mult1}. Further properties concerning
the charge multiplicity behavior as a function of energy can 
be found e.g. in reference \cite{sar}. 
Taking the $\langle \nch \rangle$ dependence on energy in GeV units
\cite{multi_review} to be $3.105\times s^{0.178}$ 
and for \RI the one given by Eq. (\ref{r1d}) one obtains
\begin{equation}
R_{\rm{1D}}\simeq 0.24 + 0.392\times \ln(\langle \nch \rangle)\ \rm{fm}.
\label{eqrvsn}
\end{equation}
\noindent
Combining the relation $d\nch/d\eta|_{\eta=0}=2.716-0.307\times 
\ln(s)+0.0267\times \ln^2(s)$ given in \cite{cms7}, where $s$ 
is in GeV$^2$ units, 
and the \RI energy dependence given by Eq. (\ref{r1d})  one has
\begin{equation}
R_{\rm{1D}} \simeq 1.08 + 0.43\times\sqrt{d\nch/d\eta|_{\eta=0}- 1.83}\ \rm{fm}.
\label{eqrvsdn}
\end{equation}
which is valid from about $d\nch/d\eta|_{\eta=0}=2$.
The \RI dependence on $\langle \nch \rangle$  
and on $d\nch/d\eta|_{\eta=0}$ are shown 
in Fig. \ref{rvsn}.
\begin{figure}[ht]
\centering{\epsfig{file=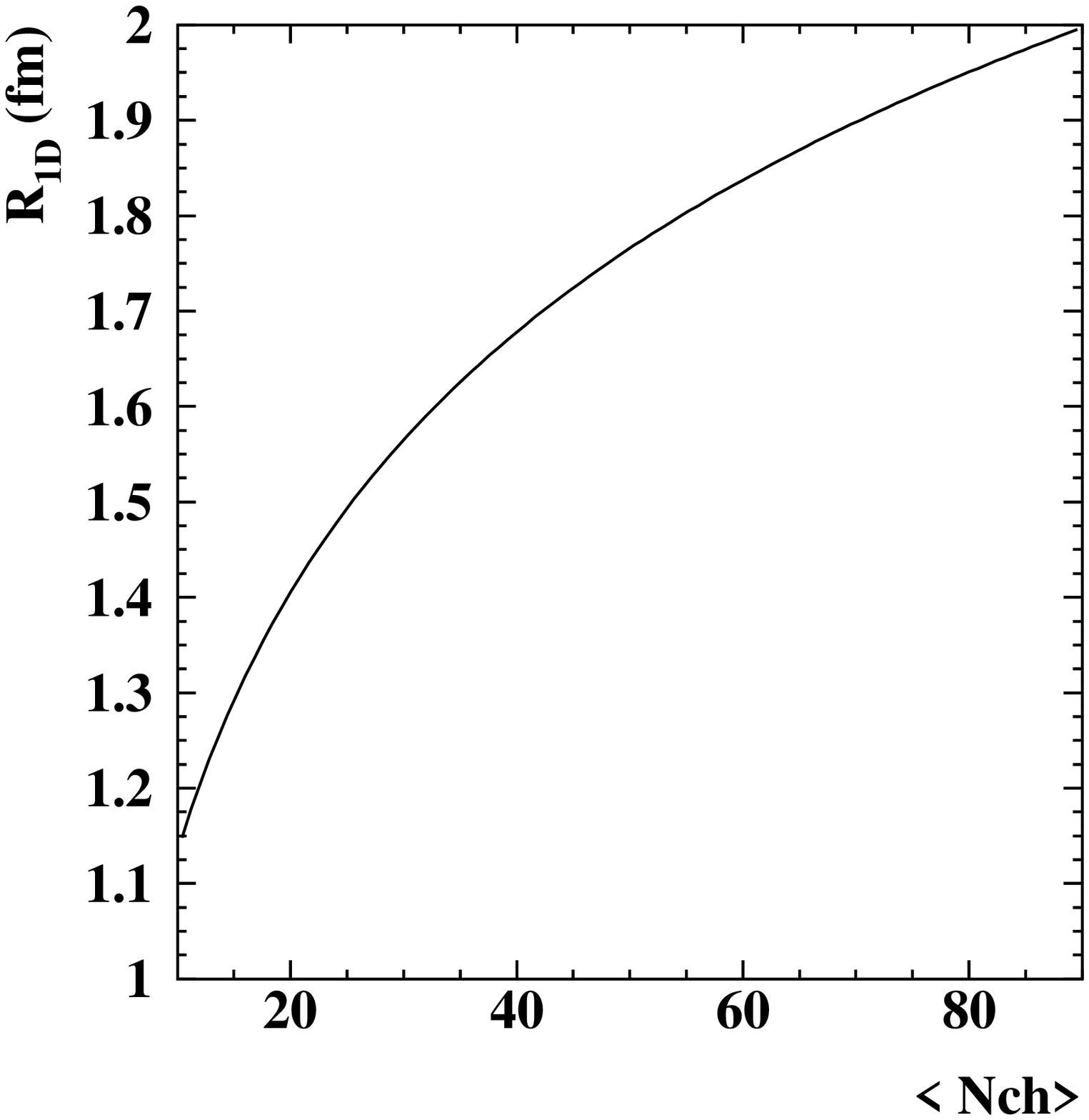,height=7.5cm,
bbllx=24pt,bblly=123pt,bburx=541pt,bbury=670pt,clip=}\ \ 
\epsfig{file=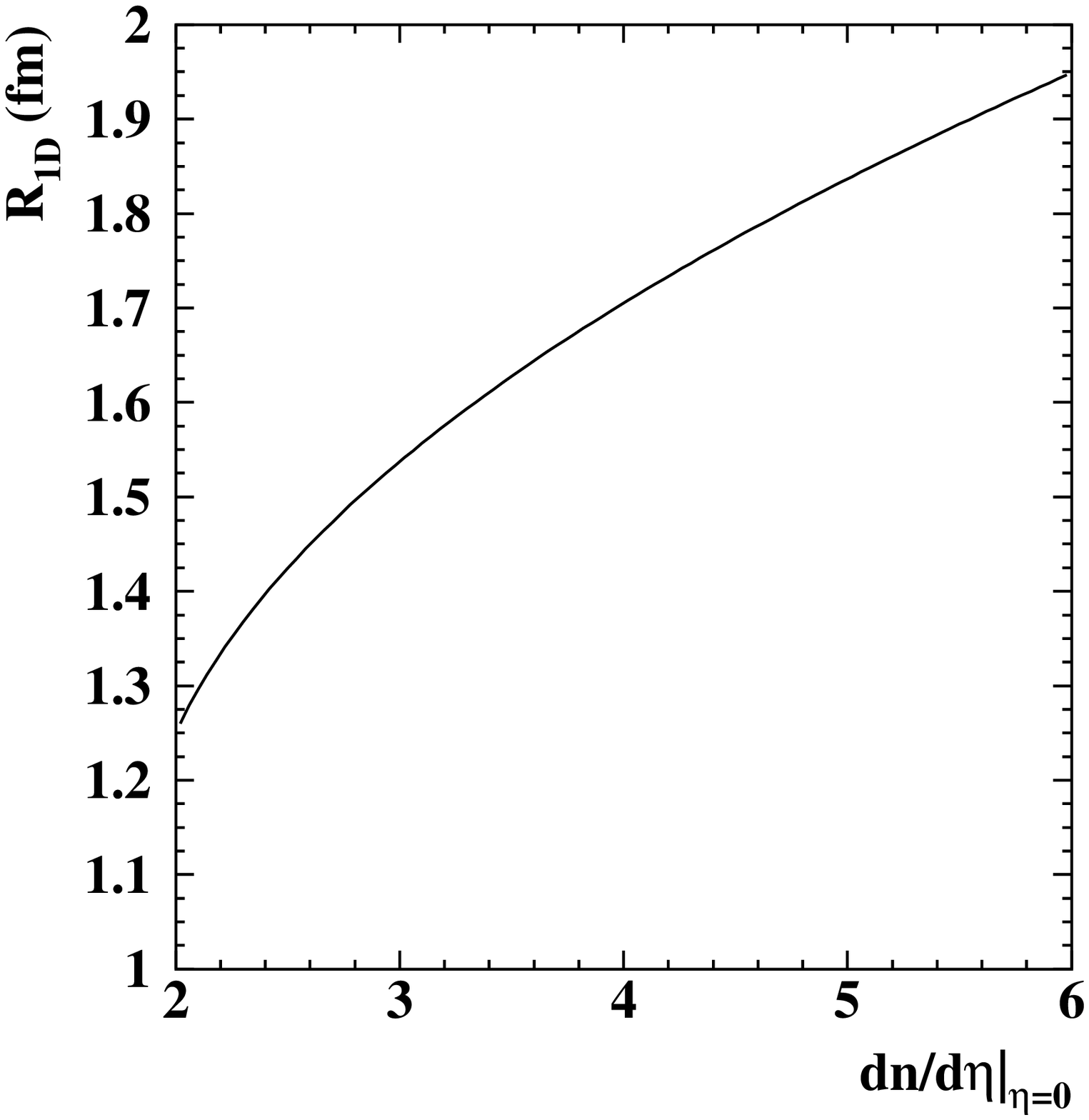,height=7.5cm,
bbllx=20pt,bblly=132pt,bburx=537pt,bbury=673pt,clip=}}
\caption{\small Left: The \RI dependence on the average
charge multiplicity $\langle \nch \rangle$ 
according to Eq. (\ref{eqrvsn}); Right: The \RI dependence
on $d\nch/d\eta|_{\eta=0}$ according to Eq. (\ref{eqrvsdn}). 
}
\label{rvsn}
\end{figure}
From this figure one may of course also estimate from the measured \RI
the values of $\langle \nch \rangle$ and 
$d\nch/d\eta|_{\eta=0}$ at the same $pp$ energy.

\section{The \RI dependence on the number of jets}
Even though \RI is seen to rise with $\langle \nch \rangle$, 
the BEC studies of the $Z^0$ hadronic decays 
at the LEP $e^+e^-$ collider have shown that the fundamental
reason for the \RI rise is the increase of
the number of the hadronic jets, $N_{\rm{jet}}$. 
An example of this feature is illustrated in Fig. \ref{opal_mult}
where \RI and $\lambda$ results obtained from
BEC analyzes of the $Z^0$ hadronic decay carried out by 
the OPAL collaboration at LEP \cite{opal_mult3} are plotted.
In the left part of the figure \RI is seen to increase with \Nch 
while $\lambda$ decreases with it. 
In the right part of this figure one observes that the \RI values
are consistently
higher for 3-jet events than those derived from the 2-jet events
while both sets of the \RI values are only slightly 
dependent on \Nch . At the same time the $\lambda$ values 
do not reveal any obvious separation between the 3 and 2-jet events       
nor do they
show a distinct dependence on the charge multiplicity.
A similar result has also been obtained by the L3 Collaboration
at LEP \cite{l3jet}.
That the \RI value is indeed linked to the
number of hadronic jets has already been successfully demonstrated
in the study of the measured dependence of \RI on $\sqrt{s}$ 
in $e^+e^-$ annihilations into hadrons \cite{bec_ee_energy}. 
\begin{figure}[ht]
\centering{\epsfig{file=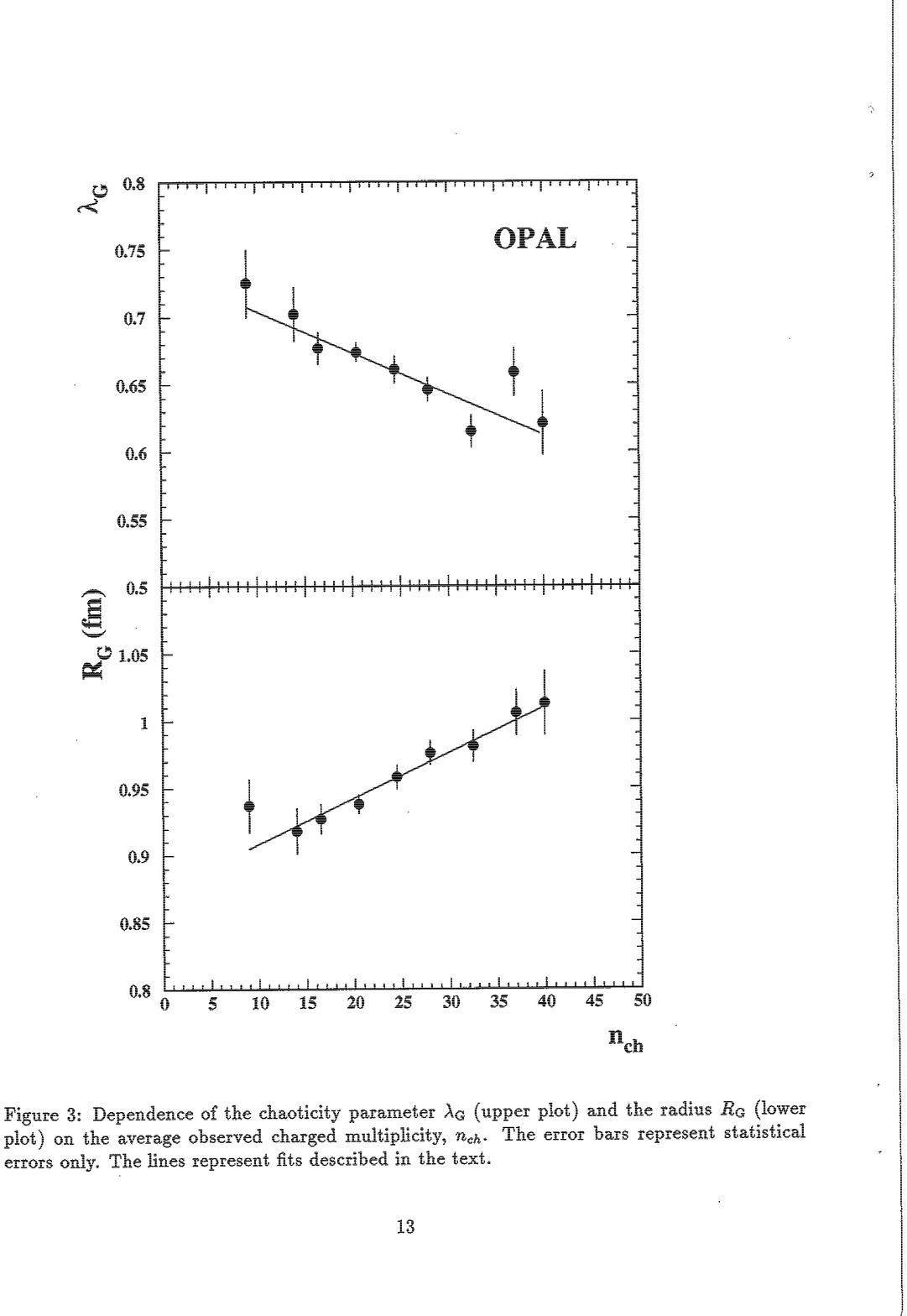,height=9cm,
bbllx=55pt,bblly=152pt,bburx=451pt,bbury=709pt,clip=}\ \ \ 
{\epsfig{file=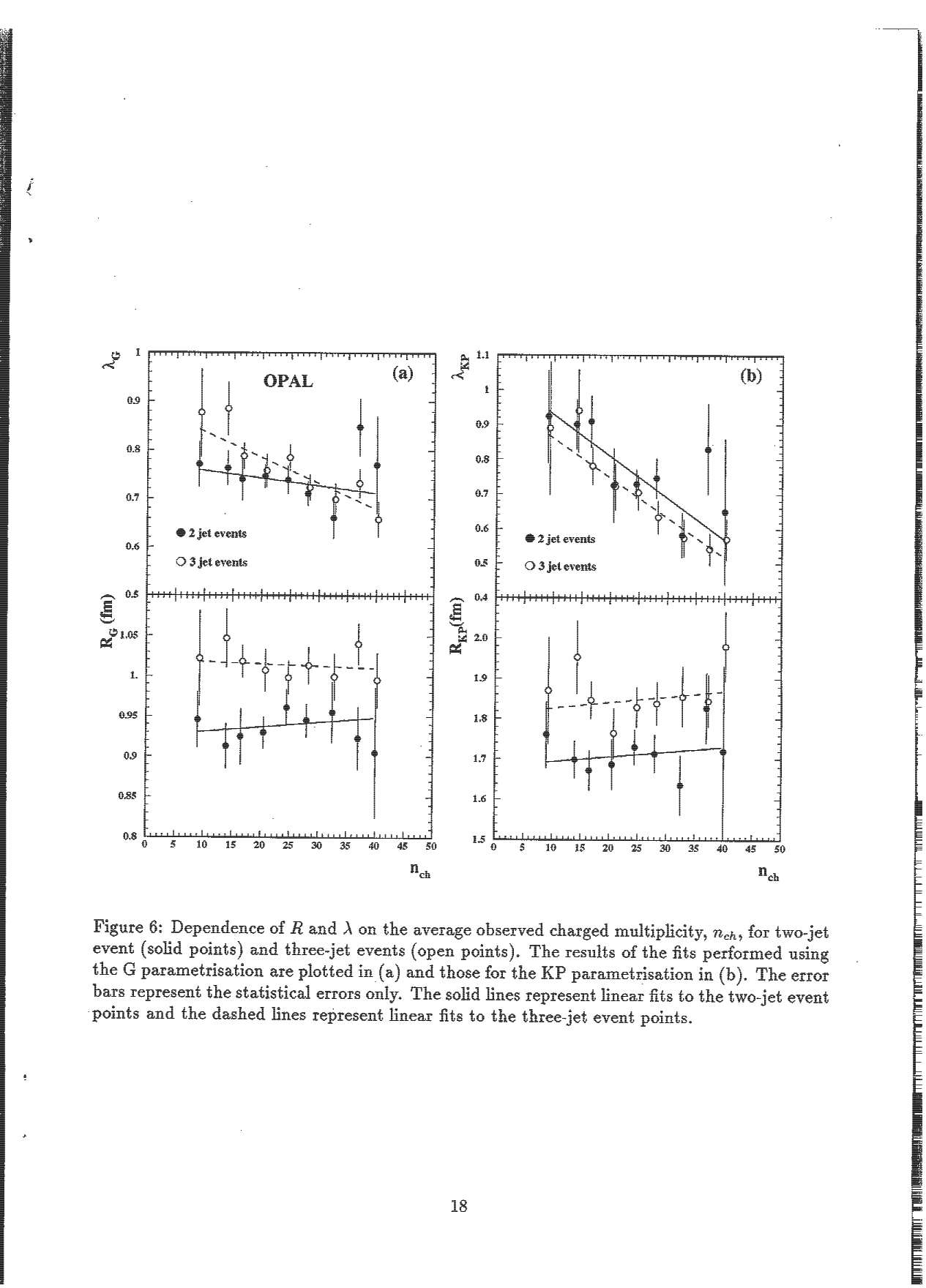,height=9cm,
bbllx=44pt,bblly=258pt,bburx=289pt,bbury=629pt,clip=}}}
\caption{\small The dependence of the chaoticity parameter
$\lambda$ and the source dimension \RI 
on the observed charge multiplicity \Nch and number of jets
in the $Z^0$ hadronic decays given in reference \cite{opal_mult3}.
Left: $\lambda$ and \RI as a function of \Nch . Right: 
The dependence of $\lambda$ and \RI on
the observed \Nch for two jet (solid points) and three
jet events (open points). The straight lines in the figures represent
linear fits to the data points. 
}
\label{opal_mult}
\end{figure}
\vspace{2mm}

\section{The mean number of jets deduced from \RI and $\lambda$}
A compilation of \RI values obtained from BEC analyzes
of identical pions 
produced in heavy ion
collisions \cite{chacon} has shown that to a good
approximation the relation $R_{\rm{1D}}=1.2\times A^{1/3}$ fm holds where $A$
is the atomic number of the colliding nucleus. This \RI dependence on
$A$ is consistent with the dependence of the radius of nuclei
on $A$ deduced from other types of experiments such as the 
electron nucleus scattering. 
The increase in \RI is attributed first 
to the rise in the number of the hadron
sources, that is the number of participating nucleon-nucleon 
collisions, as $A$ increases. The second feature which causes the 
\RI to rise is the absence of  
a BEC effect (i.e. $\lambda$=0) 
in identical boson pair correlation emerging from different sources.
This last feature has been experimentally 
verified e.g. in the process $e^+e^- \to W^+W^- \to pions$ where the
BEC effect is present in identical pion pairs emerging from
the same $W$ gauge boson but is absent when the pair was emitted
from two different $W$ gauge bosons \cite{ww}.
From this follows that whenever identical bosons are collected for a BEC
analysis from more than one source the over all correlation decreases
which results, as briefly elucidated below, in an increase in \RI .
\vspace{2mm}
  
Inasmuch that the emitting sources are similar
the \RI values obtained from BEC of two identical bosons
emitted from $S_k$ and $S_m$
sources are related by \cite{kittel,sarkisyan2,sarkisyan3} 
\begin{equation}
R_m\ =\ \frac{\lambda_m}{\lambda_k} \frac{D_k}{D_m} R_k\ ,
\end{equation} 
where $D_k$ and $D_m$  are the reduction factors corresponding 
respectively to the $S_k$ and $S_m$ sources.
It has further been 
shown that in the case that the sources have about the same 
charge multiplicity
then for $\nch \gg 1$
the reduction factor is simply given by
\begin{equation} 
D_j\  \ {\stackrel{\nch\gg 1}{\longrightarrow}}\ \ \frac{1}{S_j}\ .
\end{equation}
This inverse dependence of $D_j$ on the number of sources
holds also when the occurrence of slight deficiencies in the 
multiplicity measurement are accounted for 
while the usage of the full dilution value implies that the
system of sources fully overlap in the momentum space
\cite{kittel,sarkisyan1}.  
Thus the BEC deduced \RI values from the $S_k$ and $S_m$ sources
are related by
\begin{figure}[ht]
\centering{\epsfig{file=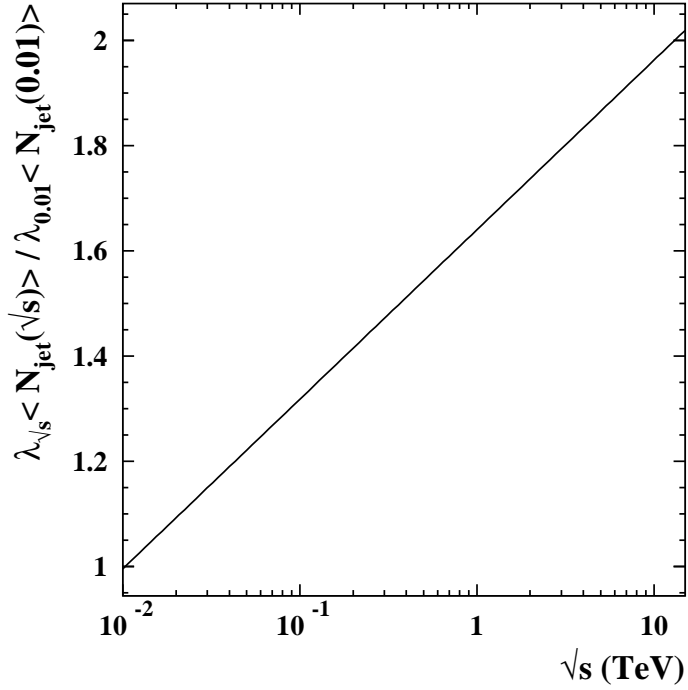,height=10cm,
bbllx=19pt,bblly=123pt,bburx=540pt,bbury=675pt,clip=}}
\caption{\small The average number of hadronic
jets produced in $pp$ collisions
multiplied by the BEC deduced $\lambda$ values 
as a function of $\sqrt{s}$ in TeV units  
obtained via Eq. (\ref{final2}) relative to
the average number of jets produced at $\sqrt{s}=0.01$ TeV 
multiplied by its $\lambda$ value.
}
\label{jetvse}
\end{figure}
\begin{equation}
R_m\ =\ \frac{\lambda_m}{\lambda_k}\frac{S_m}{S_k}R_k\ \ \ \
\Longrightarrow \ \ \ \frac{S_m}{S_k}\ =\ 
\frac{\lambda_k}{\lambda_m}\frac{R_m}{R_k}\ \ \ \
\Longrightarrow \ \ \ \ \frac{S_m}{S_k}\ =\ 
\frac{R_m/\lambda_m}{R_k/\lambda_k}\ .
\label{maineq}
\end{equation}
    
\noindent
As has been applied 
successfully in the past \cite{sarkisyan2} 
we also here, in particle reactions, associate
the hadronic jets with the boson sources.
As a consequence Eq. (\ref{maineq})
can be rewritten as 
\begin{equation}
\frac{\NJMI}{\NJKI}\ =\ 
\frac{\lambda(\sqrt{s_k})}{\lambda(\sqrt{s_m})}\times 
\frac{R(\sqrt{s_m})}{R(\sqrt{s_k})}\ .
\label{final1}
\end{equation}
Thus from \RI and $\lambda$ 
measurements via BEC analyzes at two different $pp$
energies one should be able to estimate the ratio of the 
average number of hadronic jets $\langle N_{\rm{jet}} \rangle$ 
produced over the full phase space at these two energies.
Here it is worthwhile to note the the BEC method for jet analysis 
does not depend on a
a particular choice of a jet finding algorithm and its variables.
\begin{table}[ht]
\begin{center}
\caption{\small The average number of jets
produced in $pp$ collisions at the LHC energies relative to its value at 
0.9 TeV estimated via Eqs (\ref{final1}) and (\ref{final2}).}  
\begin{tabular}{|c|c|c|c|c|c|}
 \hline
 Ref.&$\sqrt{s}$ & $R_{\rm{1D}}$ (fm)   & $\lambda$& \multicolumn{2}{c|}
{$\langle N_{\rm{jet}}(\sqrt{s})\rangle /\langle N_{\rm{jet}}(0.9)\rangle $}\\ \cline{5-6}
 & (TeV)  &Measured  & Measured&Using Eq. (\ref{final1}) &
Using Eq. (\ref{final2}) \\ \hline 
\cite{cms_09_7} &
0.90 & 1.56$\pm$ 0.12 & 0.62 $\pm$ 0.03 &1 &1  \cr\hline
\cite{cms_236} &
2.36 & 1.99$\pm$ 0.18 & 0.66 $\pm$ 0.09 &1.20$\pm$0.22&1.02$\pm$0.15 \cr
\cite{cms_09_7} &
7.00 & 1.89 $\pm$ 0.19 & 0.62 $\pm$ 0.04 &1.21$\pm$0.18&1.18$\pm$0.09 \cr
$--$ &
14.0 &\ \ [2.01 $\pm$ 0.12]* &\ \ [0.62 $\pm$ 0.03]**&
 $--$ &1.22$\pm$0.08 \cr
\hline
\end{tabular}
\label{table1}
\end{center}
\vspace{-4mm}
\hspace{0.82cm} *Extrapolated via Eq. (\ref{r1d})\ \ \ \ **Assumed  
\end{table}
\vspace{2mm}

Next we explore the possibility to evaluate the energy dependence of the 
average number of 
hadronic jets produced
in $pp$ collisions. To this end, 
the energy dependence of \RI given by Eq. (\ref{r1d})
is incorporated into Eq. (\ref{final1}) with the result that 
\begin{equation}
\frac{\NJMI}{\NJKI}\ =\
\frac{\lambda(\sqrt{s_k})}{\lambda(\sqrt{s_m})}\times 
\frac{a+b\times \ln(\sqrt{s_m})}{a+b\times ln(\sqrt{s_k})}\ .
\label{final2}
\end{equation}
Fig. \ref{jetvse} shows [$\lambda_m \times \NJM]/ [
\lambda_k \times\NJK]$  as a function
of $\sqrt{s}$ for $pp$ collisions
using in Eq. (\ref{final2})
with the values $a\ =\ 1.64$ fm and $b\ =\ 0.14$ fm and 
choosing $\NJK$   
to be the average number of hadronic jets expected at $\sqrt{s}=0.01$ 
TeV.  

\vspace{2mm}

In Table \ref{table1} we summarize the \RI and $\lambda$ 
values recently measured in $pp$ collisions at 0.9, 2.36 and 7 TeV
and their estimated values at 14 TeV, the 
originally planned LHC energy. 
Also given in the Table are the expected
$\langle N_{\rm{jet}} \rangle$ values relative 
to its value at 0.9 TeV.
As can be seen, 
the results obtained for the average number of jets   
via Eqs (\ref{final1}) and (\ref{final2}) are consistent within
errors apart from the value obtained at
$\sqrt{s}=2.36$ TeV where they differ considerable which is also
the case for its  
\RI value which is seen in Fig. \ref{rpp} to be on the higher side of
the fitted one. Another feature to notice is the
relatively slow increase of $\langle N_{\rm{jet}} \rangle$. Namely,
in an energy change by a factor of $\sim$15, between 0.9 TeV to 14 TeV,
the average number of hadronic jets increases just by about 22{\%}.

\section{Summary and conclusions}
The one dimensional \RI deduced from Bose-Einstein correlation of
identical pions produced in $pp$ collisions is seen to increase
with the center of mass energy of the interacting baryons. This 
behavior can be parametrized by $R_{\rm{1D}}=a+b\times \ln(\sqrt{s})$ fm
where $s$ is given in TeV$^2$ units and
with the fitted values of 
$a=1.64$ fm and $b=0.14$ fm. The
\RI dependence   on energy is then used to derive expressions
for its dependence on  the average charge multiplicity 
$\langle \nch \rangle$
and charge multiplicity density $d\nch/d\eta|_{\eta=0}$. 
It is further argued that the fundamental cause for
the rise of \RI with $\sqrt{s}$ is the increase with energy
of the average number of outgoing hadronic jets 
produced in $pp$ collisions. 
Two method are outlined for
the estimation of the relative average true number of
hadronic jets via BEC measurements of \RI and $\lambda$
which are applied to the recently investigated $pp$
collisions at the Large Hadron Collider  at CERN. 
An extrapolated value for the relative $\langle N_{\rm{jet}} \rangle$
value for the planned LHC $pp$ collisions at 14 TeV is also given.

\subsection*{Acknowledgments} 
Thanks are due to W. Kittel and E.K.G. Sarkisyan for many helpful 
suggestions and comments.

\end{flushleft}
\end{document}